# Black Arsenic: A Layered Semiconductor with Extreme in-plane Anisotropy

*Yabin Chen,[1,†] Chaoyu Chen,[2,†] Robert Kealhofer,[3] Huili Liu,[1,4] Zhiquan Yuan,[1] Lili Jiang,[5,6] Joonki Suh,[1] Joonsuk Park,[7] Changhyun Ko,[1] Hwan Sung Choe,[1] José Avila,[2] Mianzeng Zhong,[8,9] Zhongming Wei,[8] Jingbo Li,[8] Shushen Li,[8] Hongjun Gao,[5] Yunqi Liu,[6] James Analytis,[3,4] Qinglin Xia,[1,8,9,*] Maria C. Asensio,[2,*] Junqiao Wu[1,4,*]*

1. Department of Materials Science and Engineering, University of California, Berkeley, California 94720, United States.

2. ANTARES Beamline, Synchrotron SOLEIL, Université Paris-Saclay, L'Orme des Merisiers, Saint Aubin-BP 48, 91192 Gif sur Yvette Cedex, France.

3. Department of Physics, University of California, Berkeley, California 94720, United States.

4. Materials Sciences Division, Lawrence Berkeley National Laboratory, Berkeley, California 94720, United States.

5. Institute of Physics and University of Chinese Academy of Sciences, Chinese Academy of Sciences, Beijing 100190, China.

6. Key Laboratory of Organic Solids, Institute of Chemistry, Chinese Academy of Sciences, Beijing 100190, China.

7. Department of Materials Science and Engineering, Stanford University, Stanford, California 94305, United States.

8. State Key Laboratory of Superlattices and Microstructures, Institute of Semiconductors, Chinese Academy of Sciences, Beijing 100083, China.

9. School of Physics and Electronics, Hunan Key Laboratory for Super-microstructure and Ultrafast Process, Central South University, Changsha 410083, China.

Correspondence and requests for materials should be addressed to Q.X. (email: qlxia@csu.edu.cn), M.A. (email: maria-carmen.asensio@synchrotron-soleil.fr), and J.W. (wuj@berkeley.edu).

Two-dimensional (2D) layered materials emerge in recent years as a new platform to host novel electronic, optical or excitonic physics and develop unprecedented nanoelectronic and energy applications.[1-5] By definition, these materials are strongly anisotropic between within the basal plane and cross the plane. The structural and property anisotropies inside their basal plane,[6-9] however, are much less investigated. Herein, we report a rare chemical form of arsenic, called black-arsenic (b-As), as an extremely anisotropic layered semiconductor. We have performed systematic characterization on the structural, electronic, thermal and electrical properties of b-As single crystals, with particular focus on its anisotropies along two in-plane principle axes, armchair (AC) and zigzag (ZZ). Our analysis shows that b-As exhibits higher or comparable electronic, thermal and electric transport anisotropies between the AC and ZZ directions than any other known 2D crystals. Such extreme in-plane anisotropies are able to potentially implement novel ideas for scientific research and device applications.

When occurring at the centre of Brillouin zone, electronic band extrema (conduction band minimum (CBM) or valence band maximum (VBM)) of cubic semiconductors are mostly isotropic. Namely, their dispersion exhibits the same curvature at very small wavevectors along different lattice directions. Anisotropic bands are typically a result of anisotropic crystal structure in the long range, because in the long-wavelength limit, short-range discreteness of the lattice is averaged out. Such lattice anisotropy could lead to highly anisotropic physical properties such as electric conductivity and phonon group velocity. Meanwhile, anisotropies in physical properties may lead to unusual device applications that are not possible in isotropic materials. In the recent revival of research in black phosphorus (b-P) as a 2D semiconductor with high charge carrier mobilities, anisotropies were found in its in-plane thermal, electrical and thermoelectric transport properties,[10, 11] and anisotropy-derived applications such as polarization-sensitive photodetectors were demonstrated.[12] In this work, we report the "cousin" of b-P, black-arsenic (b-As), where more significant in-plane anisotropies are discovered. Arsenic ($3d^{10}4s^{2}4p^{3}$) generally exists as three crystalline allotropes: grey arsenic (g-As), yellow arsenic (y-As) and b-As.[13] The most stable allotrope, g-As, is metallic with buckled structure alike blue phosphorus.[14] The waxy and insulating y-As shares the similar structure with tetrahedral white phosphorus. The much less explored b-As, in contrast, has a puckered layered structure as revealed mostly in computation,[15-20] and is isostructural to the layered b-P.

The b-As crystals used in this work are natural minerals. X-ray diffraction (XRD), Auger electron spectroscopy, and energy-dispersive x-ray spectroscopy indicate a high crystallinity and purity of these crystals (Supplementary Figure S1-S2). **Figure 1**a shows the layered crystal structures of b-As with lattice parameters calculated by density functional theory (DFT). Within the puckered layer, each arsenic atom is covalently bonded to three nearest neighbours through two different types of bonds, the in-plane, longer (2.51Å) one and the out-of-plane, shorter (2.49 Å) one. These puckered layers are held together by van der Waals force separated by an interlayer distance of 5.46 Å. The layered structure allows mechanical exfoliation to prepare flakes with different number of layers. Shown in Figure. 1b is such a thin flake where a step of 0.6 nm can be seen by atomic force microscopy (AFM) scanning, consistent with the DFT predicted monolayer thickness. High-resolution transmission electron microscopy (HRTEM) image and selected-area electron diffraction (SAED) pattern in Figure 1c clearly show an orthorhombic structure with lattice spacing of 4.6 and 3.5 Å along the armchair (AC) and zigzag (ZZ) directions, respectively, in agreement with DFT-calculated values. XRD experiments (Supplementary Figure S1) further show a lattice parameter of 1.11 nm along z-direction, consistent with the DFT calculation. The impurity level of the crystal we used is lower than our detection limit, even though O. Osters, *et al.* in Ref. 16 found that pure b-As is metastable and the orthorhombic phase of $As_xP_{1-x}$ was detected up to a composition of $As_{0.87}P_{0.13}$. The lattice constants extracted from our XRD curves are a=3.64 Å, b=11.09 Å and c=4.46 Å, which quantitatively agree with the calculated values in Ref.16, a=3.74 Å, b=10.76 Å and c=4.36 Å of pure b-As. The cell volume (180.03 Å$^3$) of our sample is larger than that of $As_{0.87}P_{0.13}$ (175.79 Å$^3$), which is reasonable. In this work, very high in-plane anisotropies in the electrical transport properties (conductivity $\sigma$, mobility $\mu$), thermal conductivity $\kappa$, as well as effective mass $m^*$, were discovered in b-As crystals as summarized in Figure 1d. The anisotropy in mobility in b-As is up to a factor of ~28, much higher than any other known layered materials.

The anisotropic electronic structure of b-As is firstly proved by DFT calculation as shown in **Figure 2**a (Supplementary Figure S3): from the top centre (Z) to top edge Q (AC direction), and A' (ZZ direction). The VBM and CBM are both located at the bulk

Z point, suggesting b-As is a direct-gap semiconductor and the band gap is about 0.31 eV. At Z point, the calculated valence band effective mass is 0.35 $m_e$ and 0.56 $m_e$ along AC and ZZ direction, respectively ($m_e$ is the electron mass in vacuum). Furthermore, close to Q point, there exists one more valence valley with even lighter effective mass of 0.16 $m_e$. All these unique features are proved by the following angle-resolved photoemission spectroscopy (ARPES) measurements.

The NanoARPES[21] was employed to measure the electronic structure of b-As crystal. In Figure 2b-2c, the Fermi level of b-As lies in the band gap, resulting in the absence of spectral weight at the Fermi surface. Thus, we use constant energy contours (CECs) with intensity integrated from the binding energy window 0.35-0.40 eV to characterize its spectral intensity distribution in the 3D reciprocal space. As shown in Figure 2d-2e, the CECs consist of elliptic features at Z point and near the Q point, arising from the VBM in the $k_x$-$k_y$ and $k_y$-$k_z$ reciprocal plane. Photon-energy dependent measurements (Figure 2e) show two strong elliptic features at 99 and 149 eV, corresponding to the VBM (at the Z point) along the $k_z$ direction. In fact, CECs and valence bands in Figure 2b, 2c, and 2d are all from the bulk Z4 point. This allows us to directly compare the NanoARPES data with DFT calculation.

Comparing Figure 2c with Figure 2a, parabolic fitting to the top part of the valence band in NanoARPES yields electronic effective mass of about 0.40 $m_e$ and 0.61 $m_e$ for AC and ZZ directions at Z point, respectively. Along AC direction, the hole-like valley near Q point can be also observed with even lighter effective mass about 0.11 $m_e$. The good agreement between experimental data and theoretical calculation confirms the reliability of our DFT calculation, and highlights the in-plane anisotropies of electronic structure. Another interesting point comes from the energy position of Fermi level inside the band gap. Since DFT predicts a small bulk gap of 0.31 eV and Fermi level lies about 0.2 eV above the VBM in Figure 2c, this suggests that both electron-like and hole-like transports are accessible. Moreover, only one hole-like valley exists along ZZ direction, while along AC direction the extra hole-like valley at higher binding energy (0.4 eV) will provide extra source for charge carriers. These electronic characteristics will manifest themselves in the transport behaviour in terms of charge carrier type and mobility.

Strong anisotropies in physical properties are expected from the strongly anisotropic lattice and electronic structures. **Figure 3**a shows the calculated phonon dispersion of bulk b-As with nine optical and three acoustic branches. The anisotropic vibrational properties of b-As were investigated by angle-resolved polarized Raman spectroscopy (ARPRS)[22] with a linearly polarized incident laser (Figure 3b). At any given polarization angle, the Raman spectrum of b-As in Figure 3c has three strong peaks located at 223.6, 230.2, and 257.9 cm$^{-1}$, quantitatively consistent with the calculated frequencies in Figure 3a. By comparing to the well-indexed Raman spectrum of b-P,[22] these three peaks are attributed to the out-of-plane $A_g^1$, and in-plane $B_{2g}$ and $A_g^2$ modes, respectively. As shown in Figure 3d-e, intensity of these Raman peaks varies as the laser polarization direction rotates, forming a two-fold symmetry for $A_g^1$ and four-fold symmetry for $B_{2g}$. As is well known in the b-P-like structure,[23] the $A_g^1$ intensity is expected to vary as $(1+b\sin^2\theta)^2$, where $b$ is a parameter related to a tensor element of the vibration mode, and the B$_{2g}$ intensity to vary as $e^2\sin^2(2\theta)$, where $e$ is the Euler's number ($\theta$ is the angle between the laser polarization and AC direction of crystal). From Figure 3d-e, the AC direction is identified as the direction where the A$_g^1$ peak is the strongest and the B$_{2g}$ peak is the weakest. This is also consistent with electrical measurements as discussed below.

Phononic thermal transport of b-As nanoribbons along AC and ZZ directions were investigated to further probe the anisotropic phonon properties. Figure 3f shows a representative scanning electron microscopic (SEM) image of the device with two suspended pads bridged with a b-As nanoribbon. These b-As nanoribbons were lithographically made from exfoliated b-As flakes.[10] Thermal conductivity $\kappa$ was measured by transporting heat from the heater to sensor pad. As shown in Figure 3g, the obtained $\kappa$ along ZZ direction is about 50% higher than that along AC direction over a wide temperature range. Akin to the similar behaviour in b-P,[10] this is attributed to the anisotropic lattice properties where the lattice along ZZ direction is stiffer and less anharmonic. Compared to b-P, $\kappa$ of b-As with the similar thickness is reduced due to its higher atomic mass. The $\kappa_{ZZ}/\kappa_{AC}$ ratio of b-As reaches 1.6 at room temperature, comparable to 1.8 of b-P.[10, 24] The average group velocity ratio (calculated $v^2_{ZZ}/v^2_{AC}$ = 2.0) dominates the anisotropy in $\kappa$ of b-As (Supplementary Table S1).

The direction-dependent in-plane electrical conduction was measured from a twelve-terminal device of b-As, as shown in Figure 3b. The angle interval between two adjacent electrodes is 30º, and the six pairs of opposite electrodes have the same channel length. The conductance of these two-terminal devices measured at room temperature is shown as a function of angle in **Figure 4**a. For an anisotropic semiconductor with two-fold symmetry like b-P,[23] the conductance $S$ is expected to vary with $\theta$ as $S_\theta = S_{max}\cos^2\theta + S_{min}\sin^2\theta$, where $S_{max}$ and $S_{min}$ are conductance along the maximal and minimum directions, respectively, and $\theta$ is the angle of current flow with respect to the direction of maximum conductance. Fitting to the experimental data in Figure 4a results in $S_{max}$ = 24.3 µS along AC direction and $S_{min}$ = 3.8 µS along ZZ direction. In practice, however, there are many factors that could results in deviations from this dependence, such as fluctuation in contact resistance and contact area of these two-terminal devices, as well as error in the angle measurement. Therefore, the conductance of b-As along AC direction is 6.4 times higher than that along ZZ direction, which is qualitatively consistent with the ARPES results in Figure 2, considering that a smaller effective mass corresponds to a higher mobility.

To further elucidate the anisotropy in b-As, Hall effect measurements were carried out along ZZ and AC directions. Two types of Hall-bar devices were fabricated on two multilayer stripes lithographically fabricated from the same b-As flake (Inset of Figure 4b). Their crystalline orientation was identified by both ARPRS and electrical measurements. As shown in Figure 4b, as temperature decreases, the four-terminal longitudinal resistance $R_{xx}$ in the ZZ direction gradually increases, which confirms the semiconducting nature of b-As as indicated by both DFT calculations and optical measurements (Supplementary Figure S4). Interestingly, $R_{xx}$ in the AC direction shows a nearly temperature-independent behaviour. In Figure 4d, Hall conductivity of AC and ZZ shows opposite signs, suggesting opposite charge of the dominant carriers along the two directions. Considering the narrow bandgap $E_g \sim 0.3$ eV for bulk/multilayer b-As, free carriers can be thermally excited at room temperature, leading to coexistence of both free electrons and holes at considerable densities. In this context, a two-carrier model[25] was utilized to analyse the Hall results. The longitudinal and transverse conductivity were given by $\sigma_{xx} = \rho_{xx}/(\rho_{xx}^2 + \rho_{xy}^2)$ and $\sigma_{xy} = \rho_{xy}/(\rho_{xx}^2 + \rho_{xy}^2)$, respectively. As shown in Figure 4c-d, the concentration $n$ and mobility $\mu$ of both electrons ($e$) and holes ($h$) can be independently extracted by fitting to the magnetic field ($B$) dependent $\sigma_{xx}$ and $\sigma_{xy}$ with the equations[25] of $\sigma_{xx} = \frac{n_h e \mu_h}{1+\mu_h^2 B^2} + \frac{n_e e \mu_e}{1+\mu_e^2 B^2}$ and $\sigma_{xy} = \frac{n_h e \mu_h^2 B}{1+\mu_h^2 B^2} - \frac{n_e e \mu_e^2 B}{1+\mu_e^2 B^2}$, where $e$ is the elementary charge. The best-fit results are $n_e = 2.0\times10^{16}$ cm$^{-3}$ and $n_h = 5.3\times10^{15}$ cm$^{-3}$. Along ZZ and AC directions, the electron and hole mobility exhibit a high anisotropy as $\mu_e^{ZZ} = 376.7$ cm$^2$/Vs, $\mu_e^{AC} = 1.5$ cm$^2$/Vs, $\mu_h^{ZZ} = 60.7$ cm$^2$/Vs, and $\mu_h^{AC} = 10,606$ cm$^2$/Vs. These strong anisotropies are qualitatively consistent with the discussion based on NanoARPES and DFT analysis.

In Figure1d, the in-plane anisotropies of b-As in electrical conductance, carrier mobility, thermal conductivity and effective mass are compared with other representative 2D materials, including b-P,[10, 22, 23] graphene,[26] ZrTe$_5$,[27] SnSe,[28] and ReS$_2$.[8, 29] It can be seen that b-As displays the maximum anisotropy in electrical conductance (6.4) and carrier mobility (~28), much higher than that of all other 2D materials, including the isostructural b-P. Similar to b-P, b-As exhibits an opposite anisotropy in electrical and thermal conductivity, namely, the electrical conductivity is higher in the AC direction than ZZ direction, while thermal conductivity behaves the opposite. This may provide a material platform to implement novel device ideas such as transverse thermoelectrics.[30]

In conclusion, we discovered a new member in the family of layered semiconductors, the black As, and carried out investigations of its physical properties in the basal plane. It is found that b-As possesses high anisotropies along AC and ZZ directions in its conductance, carrier mobility, and thermal conductivity. The mobility ratio along AC and ZZ direction is the highest among all known 2D crystals. We envision that these results provide new opportunities for layered semiconductors in nanoelectronic or thermoelectric applications.

**Experimental Section**

*Materials and characterization:* The b-As material used in this work is a natural mineral which displays a shining lustre.[31] Its impurity and quality were evaluated by Auger spectroscopy and XRD. Auger spectrum shows exclusively arsenic signals without other impurity peaks over a wide energy range. The sharp peaks in the XRD spectrum (Supplementary Figure S1) indicate good crystallinity, and the extracted inter-layer distance of 5.5Å is consistent with AFM scanning and DFT calculations. Raman and photoluminescence spectra were acquired using 488 nm laser as the excitation source. The laser power was limited to < 10 μW to avoid damaging the sample. Mechanical exfoliation of the bulk crystal onto $SiO_x$ surface leads to flakes of various thicknesses. Tapping-mode scanning of AFM (Nanoscope 3D) was used to determine the thickness of the exfoliated b-As flakes. Spherical aberration-corrected HRTEM (FEI Titan 80-300 TEM) was used to characterize the atomic structure of b-As with an accelerating voltage of 80 kV. The b-As flake was transferred onto the TEM grid using a PDMS stamp.

*NanoARPES:* Due to the relatively strong interlayer bonding, vacuum-cleaved b-As surface consists of tiny randomly orientated crystalline domains, making it impossible for traditional ARPES to measure the intrinsic band structure. We performed the state-of-art NanoARPES measurements at the ANTARES beamline of the SOLEIL synchrotron facility in France. Single crystals of b-As were cleaved at room temperature and measured by NanoARPES at 90 K in ultrahigh vacuum ($1\times10^{-10}$ mbar). The photoelectron spectra were acquired by using a hemispherical analyzer Scienta R4000.

The Fermi level expansion at 90 K was measured as 32 meV, corresponding to a practical resolution of about 12 meV. The detailed NanoARPES measurement and data analysis are explained in Ref.[32] We use an inner potential 6.5 eV for the photon energy dependent analysis. Fermi level position was measured on polycrystalline Au.

*Device fabrication and transport measurement:* Electrical devices were fabricated by the standard electron-beam lithography and lift-off process. The Hall-bar devices were patterned by $SF_6/O_2$ etching of a b-As flake mechanically exfoliated onto a $SiO_x$ wafer. The electrode layer of Ti/Au (5/150 nm) were deposited by electron-beam evaporation. The devices were then annealed at 200 °C for 3 hours in argon to improve the electrical contact between the electrode and b-As flake. Hall measurements were performed using a physical property measurement system with a variable temperature and magnetic fields. Thermal conductivity was measured by using suspended pad micro-devices under the condition of steady state heat flow.

*DFT calculations:* First-principles calculations were carried out in the framework of DFT within the Perdew-Burke-Ernzerh of generalized gradient approximation and the projector augmented wave formalism implemented in the Vienna ab initio Simulation Package (VASP).[33] A $10 \times 8 \times 4$ *k*- meshes were adopted to sample the first Brillouin zone, and the cutoff energy for the plane-wave basis was set to 500 eV for all calculations. In optimizing the system geometry, van der Waals interactions were considered by the vdW-DF level with the optB88 exchange functional.[34] Both the lattice parameters and the positions of all atoms were relaxed until the force is less than 1 meV/Å. Band structures were calculated by the hybrid functional (HSE06)[35] method based on the atomic structures obtained from the full optimization. To calculate phonon dispersion, a 4 × 4 × 1 supercell was constructed. The harmonic second order interatomic force constants (IFCs) were obtained within the linear response framework by employing the density functional perturbation theory[36] as implemented in the VASP code. We get the phonon dispersion of bulk b-As using the Phonopy package.[37]

**Supporting Information**

Supporting Information is available from the Wiley Online Library or from the author.

**Acknowledgements**

This work was supported by the National Science Foundation under grant number DMR-1708448. Q.X. acknowledges funding from the China Scholarship Council (CSC) (No. 201306375019), National Natural Science Foundation of China (No. 11674400 and No. 61775241), Natural Science Foundation of Hu-nan Province of China (No. 2018JJ2511), and Innovation-driven Project of CSU (Grant No: 2017XC019). The Synchrotron SOLEIL is supported by the Centre National de la Recherche Scientique (CNRS) and the Commissariat à l'Energie Atomique et aux Energies Alternatives (CEA), France. J.L. and Z.W. acknowledges funding from National Natural Science Foundation of China (No. 61622406 and No. 11674310). Q.X. appreciates Mr. Z.H. Liu from Guangdong Longde Group Ltd. (China) for providing the black arsenic crystals. We thank Mr. M. Amani and Dr. Z. Shi for optical measurements, Dr. Y. Lee for discussion on Hall measurements, and Dr. P. F. Yang for the discussion on thermal transport. Q.X. and Y.C. thank Prof. Y.Z. Nie, Dr. K. Ding, and Dr. C. Wan for the insightful discussions on DFT calculations.

Y.C. and C.C. contributed equally to this work.

Received: ((will be filled in by the editorial staff))
Revised: ((will be filled in by the editorial staff))
Published online: ((will be filled in by the editorial staff))**References**

[1] K. S. Novoselov, A. Mishchenko, A. Carvalho, A. H. C. Neto, *Science* **2016**, *353*, aac9439.

[2] S. Das, J. A. Robinson, M. Dubey, H. Terrones, M. Terrones, *Annu. Rev. Mater. Res.* **2015**, *45*, 1.

[3] S. Manzeli, D. Ovchinnikov, D. Pasquier, O. V. Yazyev, A. Kis, *Nat. Rev. Mater.* **2017**, *2*, 17033.

[4] L. K. Li, Y. J. Yu, G. J. Ye, Q. Q. Ge, X. D. Ou, H. Wu, D. L. Feng, X. H. Chen, Y. B. Zhang, *Nat. Nanotechnol.* **2014**, *9*, 372.

[5] F. N. Xia, H. Wang, D. Xiao, M. Dubey, A. Ramasubramaniam, *Nat. Photon.* **2014**, *8*, 899.

[6] X. Ling, H. Wang, S. X. Huang, F. N. Xia, M. S. Dresselhaus, *Proc. Natl Acad. Sci. USA* **2015**, *112*, 4523.

[7] R. X. Fei, A. Faghaninia, R. Soklaski, J. A. Yan, C. Lo, L. Yang, *Nano Lett.* **2014**, *14*, 6393.14

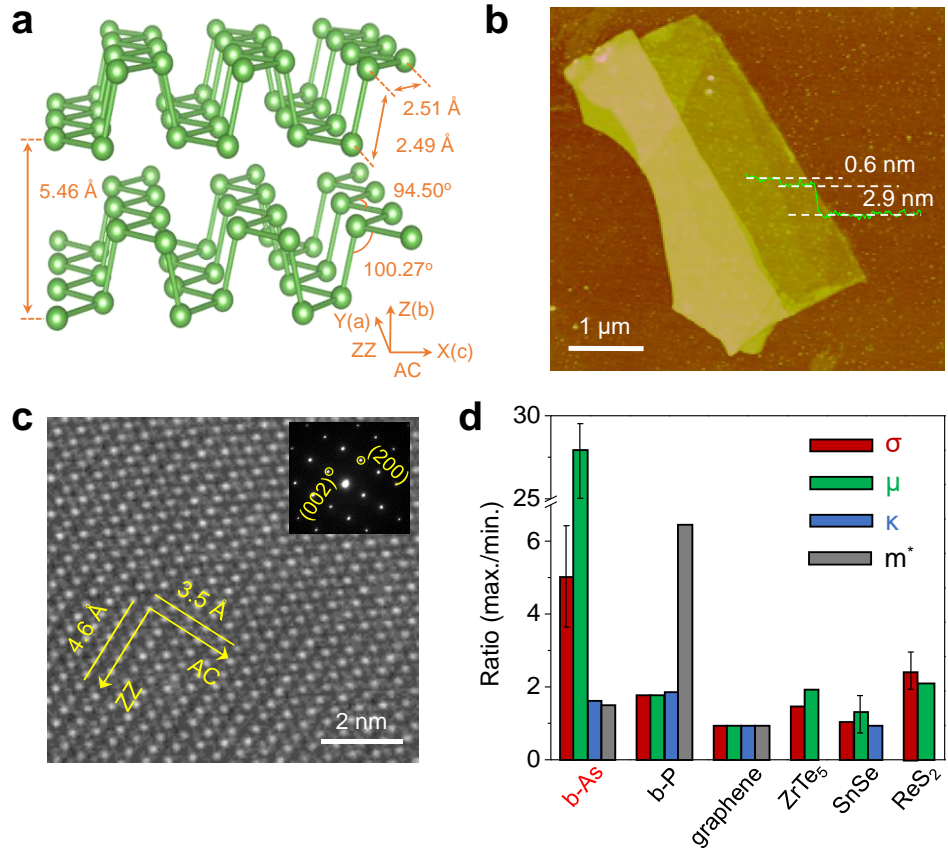

**Figure 1.** Anisotropic lattice and properties of layered b-As. a) Three-dimensional representation of b-As structure with the lattice parameters calculated by DFT. The distance between two adjacent layers is 5.46 Å. b) A b-As flake mechanically exfoliated and imaged with AFM, showing a step with monolayer thickness of ~0.6 nm. c) HRTEM image of b-As with atomic resolution. The lattice parameter of 4.6 Å and 3.5 Å is along the AC and ZZ direction, respectively. Inset is the indexed SAED pattern. d) Comparison of in-plane anisotropy in electrical conductivity $\sigma$, mobility $\mu$, thermal conductivity $\kappa$, and effective mass $m^*$ along the AC and ZZ directions of b-As and other 2D materials. Here, $m^*$ of b-P is calculated.

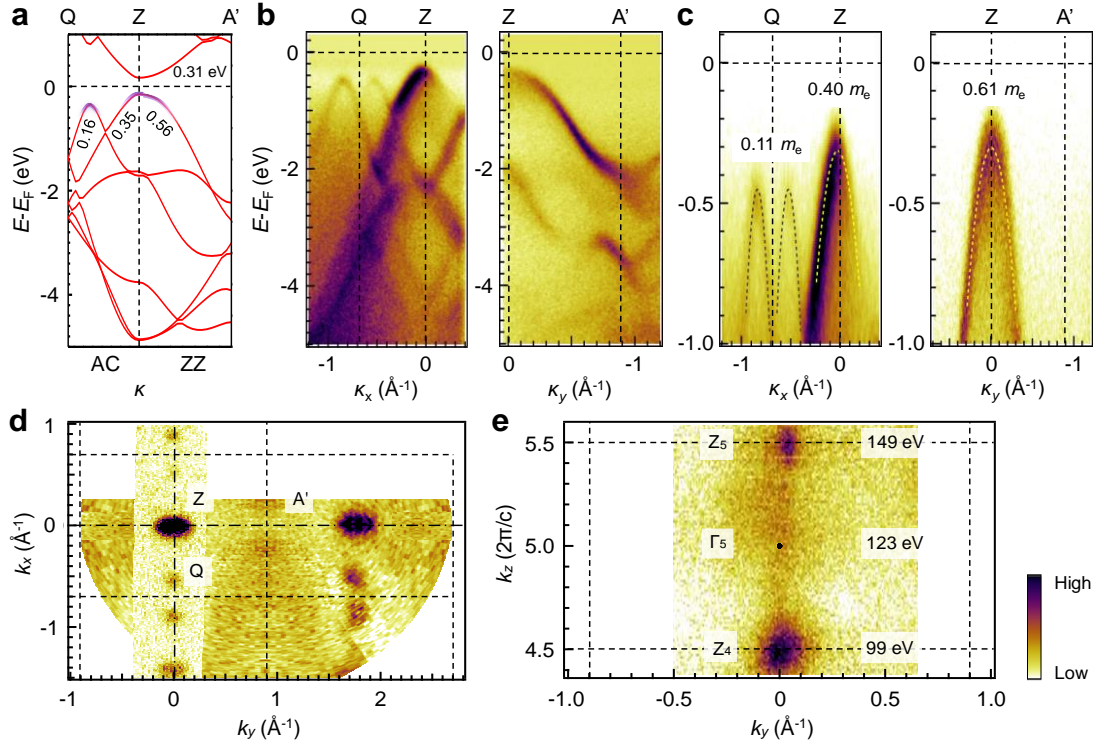

**Figure 2.** Anisotropic electronic structure of b-As obtained by DFT calculation and NanoARPES measurements. a) Band structure of bulk b-As calculated with DFT along the high symmetry directions connecting Q, Z and A' in the Brillouin zone. b) Valence band structure along QZ and ZA' direction measured by NanoARPES. c) Zoom in at the valence band structure. The dashed lines represent the momentum-distribution curves fitted dispersion with a parabolic function. The effective masses are indicated and the accuracy is about 0.02 $m_e$. d) and e) CECs of the valence bands with intensity integrated from the binding energy window 0.35 - 0.40 eV in the $k_x$-$k_y$ and $k_y$-$k_z$ reciprocal planes, respectively. NanoARPES data in b), c) and d) are measured with a photon energy of 99 eV.

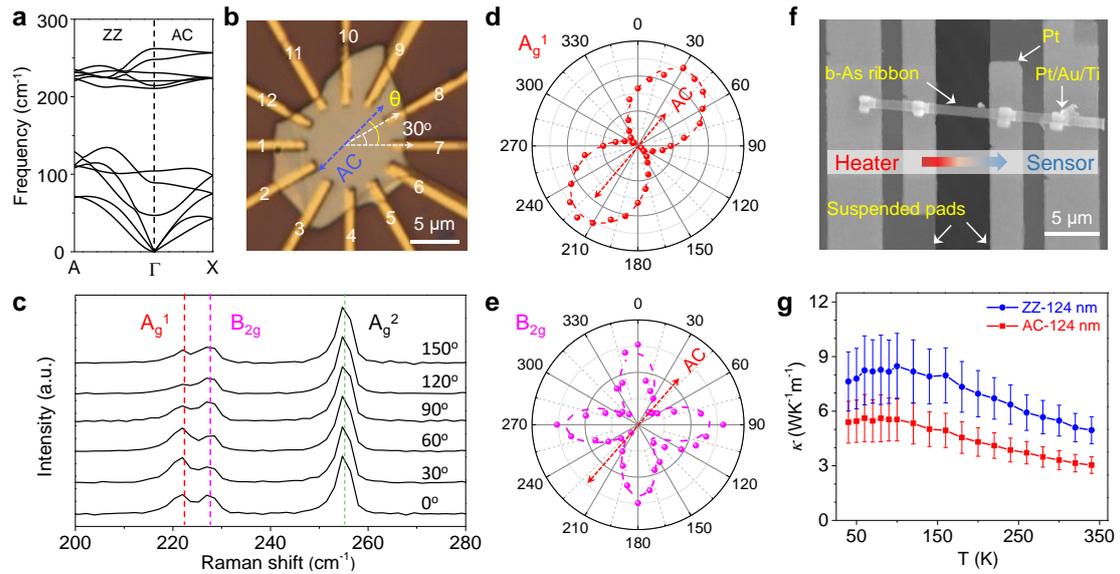

**Figure 3.** Anisotropy in phonon properties of b-As from the polarized Raman spectrum and thermal conductivity measurements. a) Calculated phonon dispersion of bulk b-As. b) Optical image of a twelve-terminal device used for Raman and electrical conductance measurements. The b-As thickness is 132 nm. c) Representative Raman spectra of the multilayer b-As with $A_g^1$, $B_{2g}$ and $A_g^2$ modes under excitation with laser polarized along different directions. d)-e) Laser-polarization dependent Raman intensity of $A^1_g$ mode d), showing two-fold symmetry with maximum along the AC direction, and $B_{2g}$ mode e), showing four-fold symmetry with minimum along the AC direction. f) SEM image of a suspended b-As device for thermal conductivity measurements. g) Temperature-dependent thermal conductivity of b-As nanoribbons along the ZZ and AC directions, respectively.

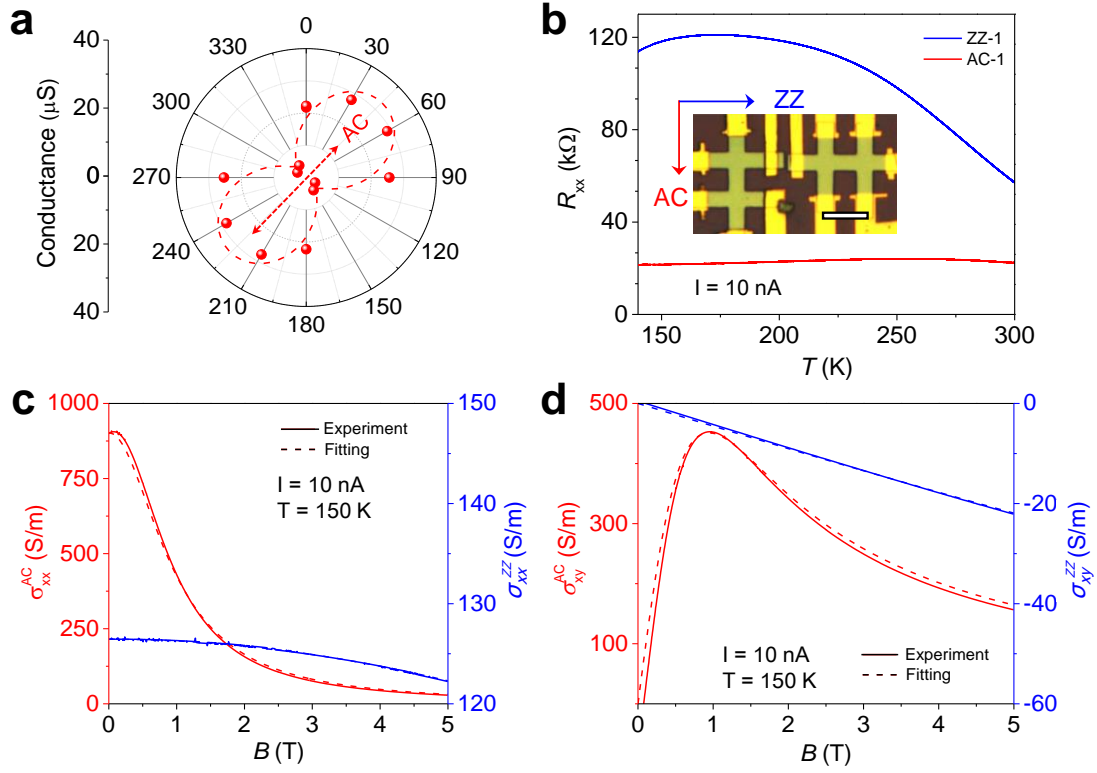

**Figure 4.** Anisotropy in electrical transport properties of b-As. a) Angular-dependent electrical conductance of multilayer b-As, showing maximum along the AC direction. b) Temperature-dependent resistance of b-As along the AC and ZZ directions. The bias current is 10 nA. Inset shows optical image of two Hall devices made of a multilayer b-As. Scale bar is 5 μm. c) Magnetic field-dependent longitudinal conductivity $\sigma_{xx}$ of the b-As measured with a current of 10 nA along AC and ZZ directions. d) Magnetic field-dependent transverse conductivity $\sigma_{xy}$ of b-As along AC and ZZ directions at 150 K. The dashed lines in a), c) and d) are the fitting curves using the two-carrier model.

**Table of content figure**

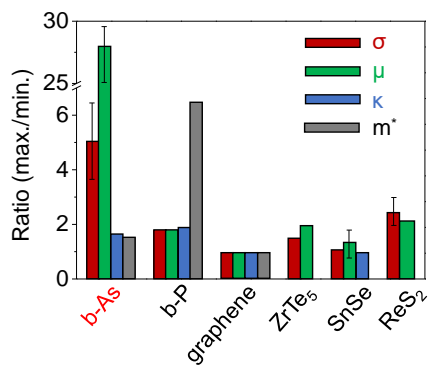

**A short summary:**

A rare chemical form of arsenic, called black-arsenic (b-As), as an extremely anisotropic layered semiconductor is reported. Systematic characterizations show that b-As exhibits higher or comparable electronic, thermal and electric transport anisotropies between armchair and zigzag directions than any other known layered crystals. Such extreme in-plane anisotropies are able to potentially implement novel ideas for scientific research and device applications.